\documentclass[%
preprint,
superscriptaddress,
 aps, prx, longbibliography
]{revtex4-2}

\usepackage{graphicx}
\usepackage{dcolumn}
\usepackage{bm}
\usepackage{amsmath}
\usepackage{upgreek}
\usepackage{amsfonts}
\usepackage{amssymb}
\usepackage[dvipsnames]{xcolor}

\usepackage{siunitx, braket} 

\usepackage[normalem]{ulem}

\begin{document}

\preprint{APS/123-QED}

\title{\textbf{A State-of-the-art Squeezed-Light Source for Hybrid Quantum Photonics: Current Performance and Remaining Challenges} 
}%
 
\author{Kai-Hong Luo}
\email{Contact author: khluo@mail.uni-paderborn.de}
\affiliation{Paderborn University, Integrated Quantum Optics, Warburger Str. 100, 33098 Paderborn, Germany}
\affiliation{Paderborn University, Institute for Photonic Quantum Systems (PhoQS), Warburger Str. 100, 33098 Paderborn, Germany}

\author{Denis Kopylov}
\affiliation{Paderborn University, Integrated Quantum Optics, Warburger Str. 100, 33098 Paderborn, Germany}
\affiliation{Paderborn University, Institute for Photonic Quantum Systems (PhoQS), Warburger Str. 100, 33098 Paderborn, Germany}

\author{Florian Lütkewitte}
\affiliation{Paderborn University, Integrated Quantum Optics, Warburger Str. 100, 33098 Paderborn, Germany}
\affiliation{Paderborn University, Institute for Photonic Quantum Systems (PhoQS), Warburger Str. 100, 33098 Paderborn, Germany}

\author{Jan-Lucas Eickmann}
\affiliation{Paderborn University, Integrated Quantum Optics, Warburger Str. 100, 33098 Paderborn, Germany}
\affiliation{Paderborn University, Institute for Photonic Quantum Systems (PhoQS), Warburger Str. 100, 33098 Paderborn, Germany}

\author{Simone Atzeni}
\affiliation{Paderborn University, Integrated Quantum Optics, Warburger Str. 100, 33098 Paderborn, Germany}
\affiliation{Paderborn University, Institute for Photonic Quantum Systems (PhoQS), Warburger Str. 100, 33098 Paderborn, Germany}

\author{Fabian Schlue}
\affiliation{Paderborn University, Integrated Quantum Optics, Warburger Str. 100, 33098 Paderborn, Germany}
\affiliation{Paderborn University, Institute for Photonic Quantum Systems (PhoQS), Warburger Str. 100, 33098 Paderborn, Germany}

\author{Benjamin Brecht}
\affiliation{Paderborn University, Integrated Quantum Optics, Warburger Str. 100, 33098 Paderborn, Germany}
\affiliation{Paderborn University, Institute for Photonic Quantum Systems (PhoQS), Warburger Str. 100, 33098 Paderborn, Germany}

\author{Torsten Meier}
\affiliation{Paderborn University, Computational Optoelectronics and Photonics, Warburger Str. 100, 33098 Paderborn, Germany}
\affiliation{Paderborn University, Institute for Photonic Quantum Systems (PhoQS), Warburger Str. 100, 33098 Paderborn, Germany}

\author{Polina Sharapova}
\affiliation{Paderborn University, Theoretical Quantum Optics, Warburger Str. 100, 33098 Paderborn, Germany}

\author{Michael Stefszky}
\email{Contact author: michael.stefszky@uni-paderborn.de}
\affiliation{Paderborn University, Integrated Quantum Optics, Warburger Str. 100, 33098 Paderborn, Germany}
\affiliation{Paderborn University, Institute for Photonic Quantum Systems (PhoQS), Warburger Str. 100, 33098 Paderborn, Germany}

\author{Christine Silberhorn}
\affiliation{Paderborn University, Integrated Quantum Optics, Warburger Str. 100, 33098 Paderborn, Germany}
\affiliation{Paderborn University, Institute for Photonic Quantum Systems (PhoQS), Warburger Str. 100, 33098 Paderborn, Germany}


\date{\today}

\begin{abstract}

Hybrid quantum photonics seeks to combine the complementary advantages of continuous- and discrete-variable quantum optics. This typically entails photon-counting measurements on entangled states generated by interfering many single-mode squeezed-vacuum (SMSV) states. However, because conventional photon-counting schemes are mode-insensitive, it is critical that the SMSV states occupy a single, well-defined mode. Achieving this requires careful engineering of the process, which determines both the spatial and spectro-temporal properties of the generated state. In addition, the ideal source must be massively scalable, capable of efficiently generating strong squeezing, and remain compatible with existing detection schemes and fiber networks. Although many platforms address one or more of these requirements, satisfying all of them simultaneously remains challenging.

Here, we present a source that simultaneously addresses all of these requirements: a single-pass, periodically poled, Type-II potassium titanyl phosphate (KTP) waveguide optimized for scalable hybrid quantum-photonic architectures. Its key defining feature is near-perfect spatial and spectral indistinguishability between the polarization-nondegenerate signal and idler fields. This design enables near-perfect interference between these two modes, resulting in picosecond-duration SMSV pulses in a single spectro-temporal mode -- the state has a measured effective mode number of $K=1.24\pm 0.3$. Furthermore, the source is extremely bright (producing up to 40 000 photons per pulse) and operates at a central wavelength of $1546~\mathrm{nm}$, optimized for fiber-network compatibility and which, in combination with picosecond duration, also enables intrinsic photon-number resolution in superconducting nanowire single-photon detectors. Although this source constitutes an ideal source in a simplified picture, the ultimate limitations of any source will be governed by complex dynamics that arise when the system is driven at high-gain or due to unavoidable loss during state generation. We have therefore developed a complete theoretical framework that enables a comprehensive photon-counting-based characterization of the source.
Our analysis, which is in strong agreement with the presented measurements, verifies the performance of our source, identifies the key factors limiting current performance, and, furthermore, exposes critical limitations of commonly used experimental methods in characterization. These simulations indicate that the platform can produce squeezing levels approaching $-20~\mathrm{dB}$ and also reveal that our detailed theoretical approach is required when producing squeezing above $-8~\mathrm{dB}$. This work establishes periodically poled KTP waveguides as a suitable platform for scalable hybrid quantum networks, while identifying key avenues for performance enhancement and elucidating the fundamental limitations governing their operation.

\end{abstract}

\maketitle
\section{Introduction}

Parametric down-conversion (PDC) sources are pivotal for photonic quantum technologies: they can be used to herald single photons in discrete-variable (DV) \cite{Alexander2025,Prevedel2007,Maring2024,Pegoraro2026} applications and to generate squeezed light for continuous-variable (CV) quantum optics \cite{slusher1985observation,wu1986generation}. Squeezed light is a fundamental building block for CV quantum optics which can be used to generate entanglement and implement CV quantum computing protocols \cite{Menicucci2006,Yokoyama2013,Larsen2019,Konno2024,GarciaBeni2025,yokoyama2026,AbuGhanem2026}. However, squeezed states are Gaussian states-- i.e. their Wigner functions have a Gaussian form---and are therefore, alone, an insufficient resource for quantum computation. Non-Gaussianity is necessary for operation beyond classical limitations and can be introduced into the system in several ways, such as feedforward \cite{Prevedel2007,Yamashima2025} or Kerr nonlinearities \cite{Rasputnyi2026}. Alternatively, one can introduce commonly used DV operations in the form of photon subtraction \cite{Wenger2004,endo2023,endo2025} or photon counting as the source of non-Gaussianity, resulting in what is typically referred to as a hybrid CV/DV system, or hybrid quantum photonic system.

From the myriad of non-Gaussian operations that can be implemented, photon counting (both click detection and photon-number resolved (PNR) detection) stands as one of the simplest, most scalable schemes. PNR detection is used as the non-Gaussian resource, for example, in Gaussian boson sampling (GBS), consisting of of multiple single-mode squeezed-vacuum (SMSV) states interfering in a passive linear-optical network -- both Gaussian operations -- before undergoing PNR detection \cite{hamilton2017gaussian,Paesani2019,zhong2020quantum,madsen2022quantum,Yu2023}. In this context, single-mode signifies that signal and idler photons of the PDC process are both generated in the same field mode. 

Although PNR detection provides the scalable, non-Gaussian resource in GBS, it also places strict requirements on the squeezed-light source because, in contrast to homodyne detection, it generally cannot resolve the spatial and spectro-temporal mode structure of the state \cite{christ2011probing,kopylov2026}. Consequently, photon-number detection effectively integrates over all modes accessible to the detector, requiring a well-defined spatial and spectro-temporal mode structure to ensure that detection-induced single-photon operations act on a controlled optical mode. Engineering sources of squeezed light such that they generate SMSV states that fulfill this condition, henceforth referred to as single-mode SMSV states is paramount to enabling hybrid quantum photonics. Furthermore, recent work reveals that more complete treatments are required to characterize and understand these processes in the high-gain regime  \cite{kopylov2026}.

Although single-mode operation is essential, it represents only one of several requirements that a SMSV source must satisfy. For practically useful squeezed-light sources for hybrid applications, one generally requires \cite{Vernon2019}:

(1) \textit{Scalability}: This includes source efficiency and the ability to multiplex many sources \cite{Kaneda2019,zhong2020quantum}. Achieving this often requires a high degree of experimental stability, particularly when phase stability is required. Shorter pulse durations ultimately permit faster repetition rates, which ensure that the shot rate of experimental runs is not limited by the source itself \cite{madsen2022quantum,sonoyama2026}. 

(2) \textit{Single-mode emission}: Sources must be designed such that the generated light is produced in a \textit{single}, well-defined mode. Waveguides or resonators can be used to ensure single-spatial mode operation, but single spectro-temporal mode structure must be engineered \cite{Rohde2007,Raymer2020,Planas2026}. This mode structure must be consistent across multiple sources to enable spatial multiplexing and robust to variations in experimental parameters to ensure stability.

(3) \textit{Strong, pure squeezing}: The source must produce squeezing levels sufficient for the desired application; quantum error correction schemes, for example, set strict squeezing bounds \cite{Fukui2018,Noh2020}. While nonlinear efficiency is important, loss is critical, as it not only reduces the squeezing level but also degrades the purity of the state by mixing it with vacuum. High nonlinear efficiency is also required to ensure stable operation at strong squeezing levels.

(4) \textit{Compatibility with detection schemes and networking infrastructure}: Integration with existing fiber networks and devices imposes constraints on the operating wavelength and pulse duration, due to loss and dispersion, respectively. These requirements favor operation at telecom wavelengths, while pulse durations in the picosecond regime are required to concurrently limit the effects of fiber dispersion and enable PNR detection in superconducting nanowire single photon detectors (SNSPDs) \cite{Schapeler2024}.

Current state-of-the-art SMSV light sources highlight the performance trade-offs inherent to different platforms. Type-0 processes provide the largest nonlinearities, increasing efficiency and brightness, but generally produce highly multimode SMSV states \cite{Folge2024,Roman2024,Houde2025}. This issue may be mitigated through the use of resonators, which reshape the spectral correlations \cite{Brecht2016} and further increase the efficiency \cite{ren2026}, but at the cost of increased complexity of experimental setup, high sensitivity to both loss and system perturbations, and typically result in nanosecond pulses that may limit the experimental shot rate and exclude the use of intrinsic PNR in SNSPDs \cite{madsen2022quantum,liu2026}. Another promising approach is to use sophisticated dispersion-engineering techniques in thin-film waveguiding platforms, although this approach has not yet been experimentally demonstrated and requires tight fabrication tolerances \cite{Houde2025}.

KTP provides an alternative solution owing to its favorable dispersion properties, which lead to a nearly 45-degree phasematching angle for downconversion in the telecom regime. The type-II process in this material produces a two-mode squeezed vacuum state (TSMSV), with signal and idler fields that are polarization orthogonal and nearly perfectly decorrelated -- previously exploited for its suitability as a heralded single-photon source  \cite{eckstein2011,Harder2013}. Here, we will demonstrate that one can also operate the system under conditions that provide spectral indistinguishability between the generated signal and idler fields, a crucial requirement for realizing true SMSV states. Under these two conditions, one produces two independent single-mode SMSV states by interfering signal and idler fields on a beamsplitter \cite{Braunstein2005}. This concept has been previously demonstrated in a spatial multiplexing scheme utilizing single-pass bulk sources\cite{zhong2020quantum}, and in a waveguided, temporal multiplexing scheme \cite{zhong2020quantum}.

Although such sources have been used to produce single-mode SMSV states in the low-gain regime \cite{Dirmeier2020}, their performance has yet to be comprehensively characterized and understood -- in particular, the current and fundamental limits of SMSV state generation in these newly developed schemes remain largely unexplored. Recent work indicates that the modal structure of the states generated in various nonlinear processes is expected to be impacted by losses during state generation \cite{houde2023waveguided,Kopylov2025quantum,kopylov2025spectral,kopylov2025bipartite,Christ2013}, and that it will also vary as the system is driven into the high-gain regime. In this regime, causality during state generation becomes increasingly important and can strongly influence the properties of the output state. These effects are commonly referred to as time-ordering or spatial-ordering effects, which we collectively refer to simply as ordering throughout this work. \cite{Folge2024,Lipfert2018,Horoshko2019,thekkadath2024gain,Sharapova2020,Quesada2015,Christ2013,Wasilewski_2006,kopylov2025spectral}. Characterization techniques beyond those that are commonly used, which typically require low-gain operation and negligible loss \cite{christ2011probing}, are necessary to verify the performance of these systems for the most demanding applications \cite{kopylov2026}. Furthermore, understanding and developing techniques for mitigating these effects is critical to ensure the optimal performance of any large system into which these sources are integrated into. For example, it has been shown that the computational complexity of the sampling task defined by GBS is reduced when the SMSV states do not reside in a single well-defined spectro-temporal mode \cite{Shchesnovich2022,Bezerra2026}, highlighting the negative impact that these complex dynamics can impart.

Here, we demonstrate and verify the performance of a single-pass, periodically poled KTP waveguide that has been optimized for use in hybrid quantum photonics applications. Furthermore, we develop a multimode theory that includes ordering effects and loss during state generation to provide a comprehensive system characterization using click-detection. The presented results agree exceptionally well with theory, even in the very-high-gain regime (with mean photon numbers up to 40,000). Simulations show that the platform is capable of producing high levels of squeezing, approaching 20 dB, in nearly single-mode emission, with a measured effective mode number of $K=1.24\pm 0.3$. This work establishes a benchmark for single-mode SMSV state generation that meets the demanding needs of large-scale hybrid quantum-photonic networks; high efficiency, high squeezing, single-mode operation, and compatibility with SNSPDs and mid-scale fiber networks, while also illuminating the current and fundamental limits of standard characterization techniques and of the source itself.

\section{Theory}

\label{sec:Theory}

In commonly used descriptions of SMSV sources, a number of assumptions are typically used -- that the PDC source produces exactly one spectro-temporal mode and is pure. Under these assumptions, the lossless SMSV is expected to have a mean photon number, $\langle n\rangle$, given by $\langle n\rangle = \sinh^2(r)$, and the degree of squeezing (minimal quadrature variance) $V = e^{-2r}$, where the squeezing parameter $r$ is proportional to the pump amplitude, the nonlinear susceptibility, and the interaction length. 
It follows that the low-gain regime is identified as the region where a linear relationship between mean photon number and pump pulse energy $E_p$ is observed, $\langle n\rangle \approx r^2 \propto E_p$. 
The impact of loss on the generated squeezing level is often added to this description as a virtual beamsplitter placed after state production. The predicted maximum degree of squeezing after loss $V_{\text{SM}}$ is then given by $\text{V}_{\text{SM}} = -10 \cdot \log_{10}\left( \eta \cdot V + (1 - \eta) \right)$. In this simplified treatment, one typically assumes that the generated state experiences average losses, $\eta$, given by propagation through half the length of the sample, $L$, resulting in $\eta=e^{-\alpha( L/2)}$, where $\alpha$ is the waveguide loss (or power attenuation) coefficient. 
This simplified single-mode theory is insufficient to describe the complex dynamics involved in state generation in lossy media, particularly when this effect is combined with effects due to ordering. 

To provide a more complete description of the source, we develop a theoretical treatment that goes beyond the standard low-loss, single-mode approximations. In the undepleted-pump regime assumed here, the PDC Hamiltonian is quadratic, resulting in the generation of a Gaussian quantum state.
The PDC generation process can then be described by the multimode spatial Langevin equation~\cite{Kopylov2025quantum, kopylov2025spectral} and analyzed using the framework of Gaussian states. Our approach is able to capture the essential properties of the source, even in the very high gain regime, by including multimode structure and ordering effects.

We describe the type-II PDC
in a discrete frequency space $(\omega_1, ..., \omega_N)$. The signal and idler fields at position $z$ along the waveguide are given by vectors of bosonic operators.
Introducing the correlation matrices $\mathcal{D}_a$, $\mathcal{D}_b$ and $\mathcal{C}_{ab}$ for slowly-varying operators with elements
\begin{align*}
    [\mathcal{D}_a(z)]_{nm} &=  \braket{{\hat{a}^\dagger(z, \omega_n) \hat{a}}(z,\omega_m) } e^{i [k_s(\omega_m) - k_s(\omega_n)] z  },\\
    [\mathcal{D}_b(z)]_{nm} &=  \braket{{\hat{b}^\dagger(\omega_n, z) \hat{b}}(\omega_m, z) } e^{i [k_i(\omega_m) - k_i(\omega_n)] z  },\\
    [\mathcal{C}_{ab}(z)]_{nm} &=  \braket{{\hat{a}(\omega_n, z) \hat{b}}(\omega_m, z) } e^{i [k_s(\omega_n) + k_i(\omega_m)] z  },
\end{align*}  
where $k_s(\omega)$ and $k_i(\omega)$ are the wave vectors for the signal and idler field,
their spatial evolution is described by the spatial master equations that incorporate both losses and nonlinear coupling:
\begin{align}
    \dfrac{d \mathcal{D}_a(z) }{d z} &= i \Big[\mathcal{D}_a(z) R_a - R_a^*\mathcal{D}_a(z) \Big] + 
                i\Gamma  \Big[ J(z)  \mathcal{C}^*_{ab}(z)  - J(z) \mathcal{C}_{ab}(z) \Big],  \\
    \dfrac{d \mathcal{D}_b(z) }{d z} &= i \Big[\mathcal{D}_b(z) R_b - R_b^* \mathcal{D}_b(z) \Big] +
                i\Gamma \Big[ J^T(z)   \mathcal{C}^*_{ab}(z)  - (J^T(z))^* \mathcal{C}_{ab}(z) \Big],  \\
    \dfrac{d \mathcal{C}_{ab}(z) }{d z} &=  i \Big[ \mathcal{C}_{ab}(z)R_b + R_a \mathcal{C}_{ab}(z)   \Big] + i\Gamma \Big[ J^T(z) + J^T(z) \mathcal{D}_a(z)  + J(z) \mathcal{D}_b(z) \Big] . 
\end{align}
Here $\Gamma$ is the nonlinear coupling coefficient (parametric gain)
while the matrix $J(z)$ is the $z$-dependent coupling matrix with elements  
        \begin{equation}
         [J(z)]_{nm}  = S(\omega_n + \omega_m) e^{i\Delta k(\omega_n, \omega_m)z } ,
        \end{equation}
        where $S(\omega)$ is the pump field spectrum and the phase mismatch  $\Delta k(\omega_n, \omega_m)=k_p(\omega_n+\omega_m)-k_s(\omega_n)-k_i(\omega_m) - k_\text{{QPM}}$;  $k_p(\omega)$ is the pump wave vector, and $k_{QPM} = 2\pi/\Lambda$ is the quasi-phase-matching wave vector with poling period $\Lambda$. 

The losses for signal and idler subsystems, assumed to be frequency independent over the spectral range of interest, are included as diagonal matrices $R_a = \mathrm{diag}_N(i\alpha_s/2)$ and $R_b = \mathrm{diag}_N(i\alpha_i/2)$, respectively, where $\alpha_s$ and $\alpha_i$ are the field-amplitude loss coefficients.
The notation $[.]^*$ indicates the complex-conjugation of a matrix.

The correlation matrices $\mathcal{D}_a\equiv\mathcal{D}_a(L)$, $\mathcal{D}_b\equiv\mathcal{D}_b(L)$ and $\mathcal{C}_{ab}\equiv\mathcal{C}_{ab}(L)$ at the waveguide output provide complete information about the quantum state generated via type-II PDC. 
Using these matrices, the key spectral properties can be derived. For example, the spectral photon-number distribution for the signal field is given by the diagonal elements
\begin{equation}
    \braket{\hat{n}_a(\omega_m)} \equiv \braket{\hat{a}_m^\dagger \hat{a}_m} = [\mathcal{D}_a]_{mm},
    \label{eq_spectrum}
\end{equation}
while the total number of generated photons
\begin{equation}
   N_a = \sum_m \braket{\hat{n}_a(\omega_m)} = \mathrm{Tr}(\mathcal{D}_a). 
\end{equation}

The joint spectral intensity (JSI) is often used to characterize PDC states \cite{christ2011probing}, and in this work is defined as
\begin{equation}
    \mathrm{JSI}(\omega_n,\omega_m) \equiv \braket{\hat{n}_a(\omega_n) \hat{n}_b(\omega_m)} =
    \braket{\hat{a}^\dagger_n\hat{b}^\dagger_m}\braket{\hat{a}_n\hat{b}_m}  +
    \braket{\hat{a}^\dagger_n\hat{a}_n}\braket{\hat{b}^\dagger_m\hat{b}_m}.
    \label{eq_JSI_detailed} 
\end{equation}

A commonly used approach to analyze the mode structure of PDC light is to use the two-photon, or biphoton, state approximation and apply a Schmidt decomposition to the resulting expression \cite{Rohde2007}. This approximation neglects higher photon-number contributions. Alternatively, the Schmidt modes can be obtained from a Bloch-Messiah reduction. However, neither of these approaches is sufficient in the high-gain regime or in the presence of loss during state generation.
In these cases, the modal structure is obtained through the spectral decomposition of the correlation matrices $\mathcal{D}_{a,b}$~\cite{kopylov2025spectral}. 
Namely, their diagonalization yields $\mathcal{D}_a  = V_a  \Lambda_a  V_a^\dagger$ and $\mathcal{D}_b = V_b  \Lambda_b  V_b^\dagger$, 
where $V_{a,b}$ contain the Mercer-Wolf modes (first-order coherence basis) and $\Lambda_{a,b}$ are diagonal matrices. 
The values $\lambda^{a,b}_i \equiv [\Lambda_{a,b}]_{ii}$ represent the number of photons in the $i$-th Mercer-Wolf mode.
The effective number of modes for the subsystems $a$ and $b$ is given by
\begin{equation}
    \mu_{a,b} =   \dfrac{   \left[\sum_i  \lambda^{a,b}_i  \right]^2  }{\sum_i (\lambda^{a,b}_i)^2} .
    \label{eq_number_of_modes}
\end{equation}

Experimentally, the typical method for determining the number of modes is through the second-order auto-correlation function $g^{(2)}$ of the signal or idler fields, measured with detectors whose integration time exceeds the pulse duration \cite{christ2011probing}. 
The relationship between the $g^{(2)}_{a,b}$ and the number of modes is given by
\begin{equation}
    g^{(2)}_{a,b} = 1 + 1/\mu_{a,b}
    \label{eq_g2func_modes},
\end{equation}
providing a direct connection between the theoretical mode decomposition and experimental observables.
In our experiment, the $g^{(2)}$ is measured of the signal field is measured with click-detectors, which reveal saturation and imperfect separation of the signal and idler fields on the PBS (see next section).
The details regarding the numerical simulation of the $g^{(2)}$ measurements in a realistic setup are given in Appendix \ref{App:ExpImp}.

The theory until this point considers only the TMSV state produced by the Type-II KTP waveguide. To probe the properties of the generated SMSV state in a way that also includes experimental imperfections, we model the signal and idler modes interfering on a 50:50 beamsplitter. The two outputs of the beamsplitter may consist of many spectro-temporal modes, therefore, to determine the degree of squeezing in one of these outputs we probe well-defined measurement modes. This theory treatment corresponds experimentally to defining a particular local oscillator mode in a homodyne measurement. Furthermore, it can be shown that there exists a mode with a maximal degree of squeezing Ref.~\cite{Simon1994,Kopylov2025quantum}, which we can optimize for in order to unambiguously determine the maximum degree of squeezing present in the SMSV state generated in this way (see details in Appendix \ref{App:ExpImp}).

\begin{figure}
\centering
\includegraphics[width = 0.8\linewidth]{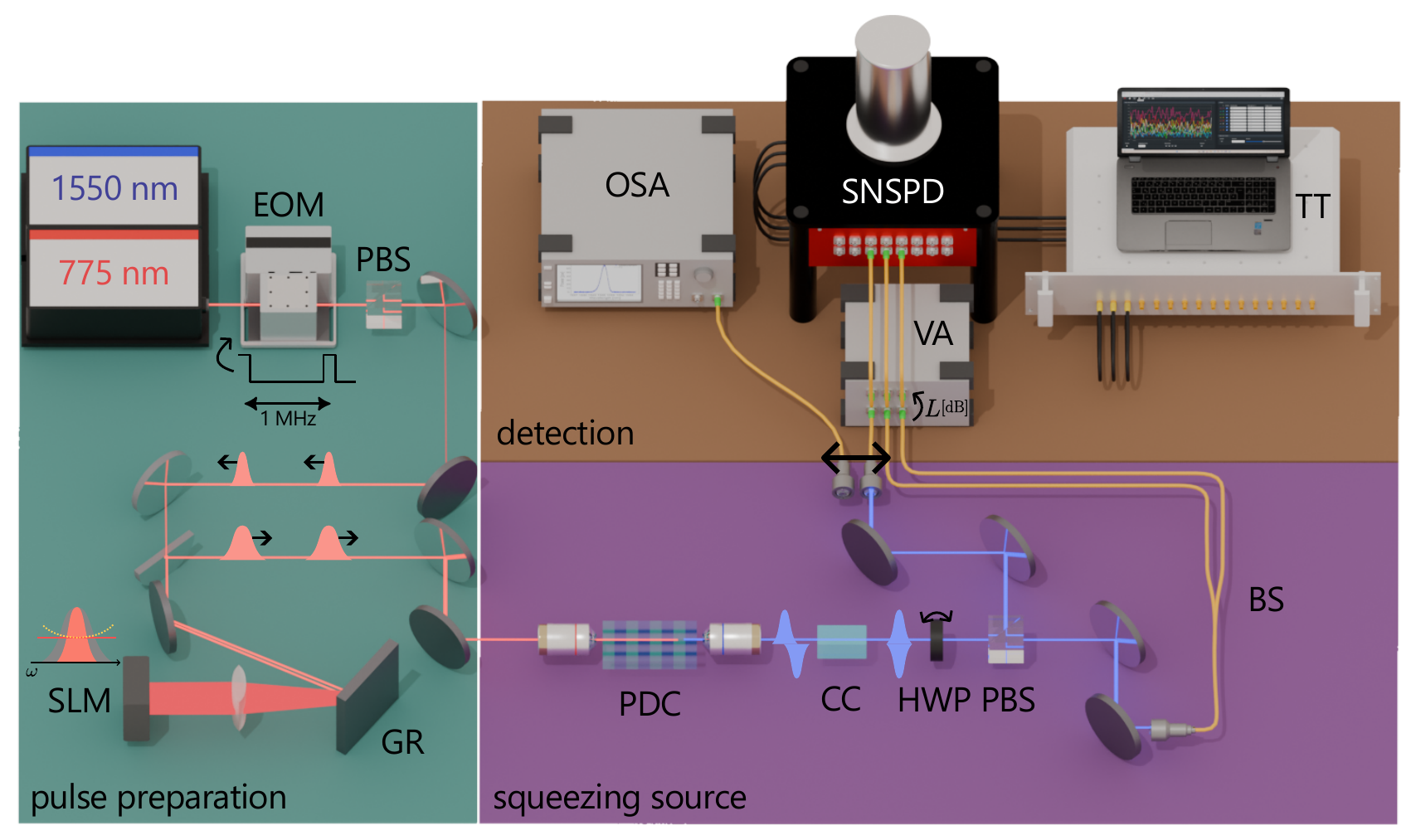}
\caption{Experimental setup. A pulsed laser with 80MHz repetition rate is first picked down to 1MHz using an electro-optic modulator (EOM) followed by a polarizing beam-splitter (PBS). The pulses are spectrally shaped using a 4f pulse shaper consisting of a grating (GR) and spatial light-modulator (SLM). The shaped pulses pass through the waveguided parametric down-conversion (PDC) source followed by a temporal compensation crystal (CC) to synchronize signal and idler fields. A half-wave plate (HWP) and polarizing beam-splitter (PBS) act as a variable beamsplitter for each field, after which both fields are fiber-coupled. The vertically polarized idler photons are either measured using an optical spectrum analyzer (OSA) or a pass through a variable attenuator (VA) before detection using an SNSPD connected to a time-tagger (TT). The horizontally polarized signal photon is split using a fiber beamsplitter (BS) and then passes through the VA and is detected using two SNSPDs. Full details provided in text.
}
\label{fig:Exp-Setup}
\end{figure}

\section{Experiment}


The presented experimental apparatus provides the foundation for the detailed characterization presented in the following sections and is designed to enable characterization of our photon-pair source in both low-gain and high-gain regimes using various measurement schemes. 
The system, shown schematically in Fig. \ref{fig:Exp-Setup}, enables click-detection measurements -- including JSI, second-order correlation  functions ($g^{(2)}$), and Hong-Ou-Mandel (HOM) interference —and high-power direct detection using an optical spectrum analyzer (OSA) (Anritsu MS9740B). 

The pump source is based on a frequency comb laser (Menlo Systems SmartComb) that produces telecom-band pulses at a repetition rate of 80 MHz. These pulses are subsequently frequency doubled (Menlo Systems M-780) to generate pump pulses centered at 772.5 nm, with a maximum bandwidth of approximately 1 nm and a maximum average power of approximately 900 mW. An electro-optic modulator (QUBIG GmbH HVOS-NIR-1.5k100) operating as a pulse picker is used to reduce the repetition rate to 1 MHz to avoid detector latching. Although the laser system is closely matched to the requirements set by the waveguide for (nearly) single spectro-temporal mode operation, fine tuning of the bandwidth and phase profile can be achieved using a programmable spatial light modulator (SLM) in a folded 4f pulse‑shaper configuration. 
In this setup, the grating and cylindrical lens map spectral components of the incoming pulses to vertical slices of the SLM which can be operated on individually \cite{weiner2011ultrafast}. The prepared pump pulses are then injected into the waveguide. 

Our waveguide source is a 20\,mm-long periodically poled KTP waveguide (AdvR Inc.) designed to produce nearly perfect spectrally indistinguishable, decorrelated type-II PDC centered around 1546nm. The waveguide features a positive phase-matching angle of approximately 57 degrees (See Appendix \ref{App:Phasematching}) and has total losses consisting of propagation losses of $0.25 \pm 0.03$\,dB/cm and a Fresnel reflection of approximately 8\% upon exiting the waveguide due to having uncoated facets.

Following the PDC process, the generated light passes through a filtering stage, consisting of a coated silicon filter providing $\sim$100 dB of pump suppression, and a 12 nm bandpass filter. This bandpass filter is much broader than the expected bandwidth of the process of only a few nanometers and is used to remove background photons -- not to shape the spectrum of the generated light. To manage the group delay introduced by the waveguide, the generated light also passes through a delay compensation crystal \cite{stefszky2025benchmarking}.

We transform the output state to a single-mode squeezed vacuum (SMSV) state using a HWP and polarizing beam splitter (PBS) configuration -- enabling flexible switching between characterization modes. 
At $\theta_\mathrm{HWP}=0^\circ$, the PBS deterministically separates signal and idler modes, allowing for JSI measurements or $g^{(2)}$ characterization of the signal field through the use of a Hanbury Brown-Twiss (HBT) setup, commonly used techniques for characterizing TMSV states \cite{christ2011probing}. 
At $\theta_\mathrm{HWP}=45^\circ$, the PBS functions as a 50:50 beam splitter (BS), resulting in the production of two independent squeezed states at the outputs of the PBS.

The two outputs from the PBS are fiber-coupled; one output includes a fiber beamsplitter before detection on two SNSPDs to implement a HBT interferometer for correlation function measurements, while the second output can be switched between an OSA for direct spectral measurements or an SNSPD for mean photon number or JSI measurements. We utilize SNSPDs (Single Quantum) with $\sim$90\% efficiency and 20 ps jitter in combination with a 2 ps resolution time tagger (Time Tagger X, Swabian Instruments).
Calibrated variable attenuators (VAs, OSICS OS-ATN-SCL-P-58) before the SNSPD prevent detector saturation.
The total system efficiency for the SNSPD measurement configuration, measured via the Klyshko method, was found to be $45\% \pm 3\%$ and $44\% \pm 2\%$ for signal and idler arms, respectively. Dispersive fiber spools used for JSI measurements showed transmissions of approximately 45\% and 50\% for the signal and idler beams, respectively.

\section{Results}

In this section, we present a comprehensive characterization of the source using complementary photon-counting and spectral measurements. Efficiency and brightness are determined by measuring the generated mean photon number in the signal field (defined as the mean number of photons exiting the source in the signal field), spectral broadening is examined using both OSA and JSI measurements, the modal structure of the generated fields is inferred using JSI measurements and correlation function measurements, and indistinguishability between signal and idler is verified using a HOM-type measurement. We find close agreement between the experimental results and our rigorous model across all of our studies, validating the high performance of our source and confirming its suitability for a wide range of quantum photonics applications. Furthermore, our rigorous modeling allows us to predict the capabilities of our source for generating high levels of squeezing and to identify the limitations that emerge as the system is driven toward these high levels of squeezing.

\begin{figure}
\centering
\includegraphics{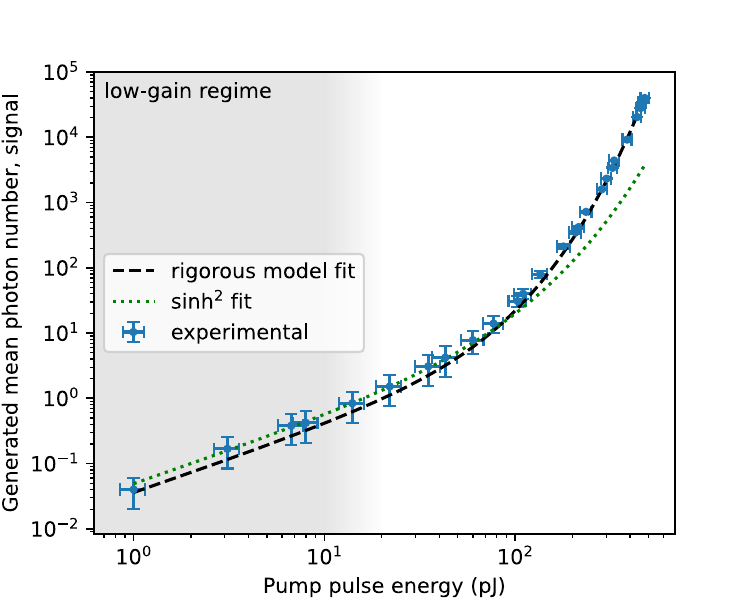}
\caption{
Generated mean photon number per pulse versus pump pulse energy. The measurement data (blue dots), with error bars representing counting error, deviate from a single-mode fit at high pump pulse energies ($\sinh^2(r)$, green dotted line). The behavior predicted by the rigorous model (black dashed line) presented here is in good agreement.
}
\label{fig:Exp-Sinh2}
\end{figure}

\subsection{Efficiency and Brightness}

The efficiency and brightness of our source, which defines its scalability, is determined by measuring the generated mean photon number of the state, the HWP behind the source (See Fig. \ref{fig:Exp-Setup}) was set to deterministically separate the signal and idler photons at the PBS and the detected photon rate of the signal field was recorded as the pump power was varied. Variable attenuators were placed in the beam path to reduce the intensity, maintaining a desired detected single-count rate —typically 100 kHz —to limit the impact of higher photon number contributions. 

The total number of generated photon pairs was determined from the detected photon number $\langle n_{\mathrm{det}} \rangle$ by correcting for both the transmission of the calibrated variable attenuator ($\eta_{\mathrm{VA}}$) and the setup efficiency quantified by the Klyshko efficiency ($\eta_K$): 
\begin{equation} \langle n_{\mathrm{gen}} \rangle = \frac{\langle n_{\mathrm{det}} \rangle} {\eta_K \eta_{\mathrm{VA}}}, \end{equation} where, the Klyshko efficiency was measured in the absence of additional attenuation.

The results of the brightness measurements are shown in Fig. \ref{fig:Exp-Sinh2} and reveal that our source is highly efficient and capable of producing states of very high brightness with generated mean photon numbers greater than 40,000 (at the highest available pump pulse energy of 480pJ). The gray-shaded region provides an approximate indication of the parameter range over which the system can be assumed to operate in the low-gain regime, where simplified theories generally provide a reasonable description of its behavior, as will be demonstrated throughout this manuscript. In the low-gain regime, the mean photon number per pulse of the signal field is 20 photons/nJ (corresponding to a source brightness of 20 pairs/nJ). It is evident from Fig.~2 that the simplified model (green dotted line) increasingly deviates from the experimental data (blue dots) as the pump pulse energy increases, whereas the rigorous model (black solid line) remains in good agreement. More specifically, the rigorous treatment (which includes both multimode and ordering effects) predicts a faster growth of the state than the simplified single-mode theory, in agreement with observations reported in previous studies \cite{christ2011probing,Vendromin2024}.

\subsection{Joint spectral intensity}

\label{Sec:JSI}

The JSI provides insight into the spectral correlations between signal and idler fields and is also commonly used to estimate the number of modes in these fields. To measure the JSI we employ dispersive-fiber spectroscopy (also known as time-of-flight spectroscopy), which maps the spectral information onto photon arrival times \cite{Avenhaus09}. Long dispersive fibers, with a dispersion of approximately $-430$~ps/nm at 1550~nm, and which are wavelength calibrated using a tunable CW laser (EXFO T200S), were incorporated into both outputs of the PBS. By recording the arrival time of each photon relative to the pump pulse, we are then able to construct a two-dimensional coincidence histogram, the JSI. 

\begin{figure}
\centering
\includegraphics{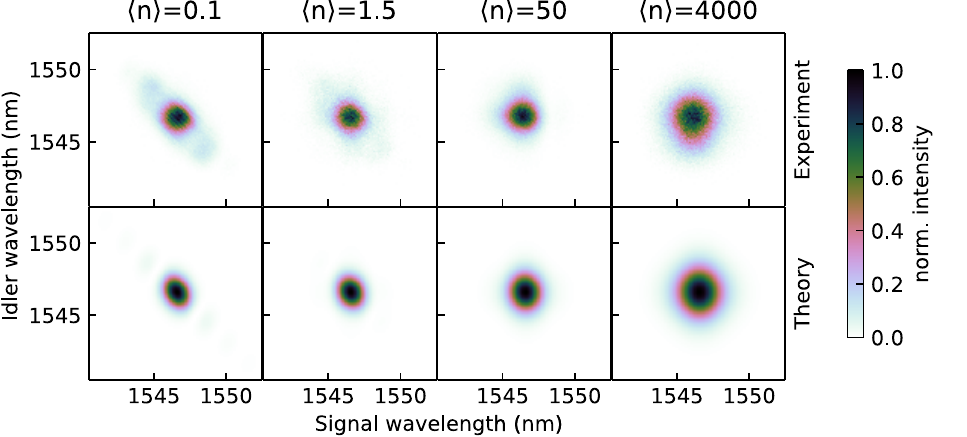}
\caption{JSI measurements and theory. At higher mean photon numbers, the phasematching side lobes diminish, the JSI broadens, and decorrelation increases.}
\label{fig:Exp-JSI}
\end{figure}

The results of the dispersive-fiber spectroscopy are presented in Fig.~\ref{fig:Exp-JSI}. Across all mean photon numbers the JSIs indicate a high degree of decorrelation and spectral indistinguishability and are in strong agreement with the behavior predicted by our theory. In the low-gain regime, the characteristic side lobes of the phase-matching function are clearly visible. As the gain increases, the side lobes diminish and a single dominant mode emerges in the JSI. Furthermore, at the highest mean photon numbers, this dominant mode undergoes significant spectral broadening. These effects are predicted by our theory and are consistent with previously reported results on high-gain operation \cite{Christ2013,thekkadath2024gain,houde2023waveguided,taheri2026}. This broadening effect is further investigated in the following section.

\begin{figure}
\centering
\includegraphics{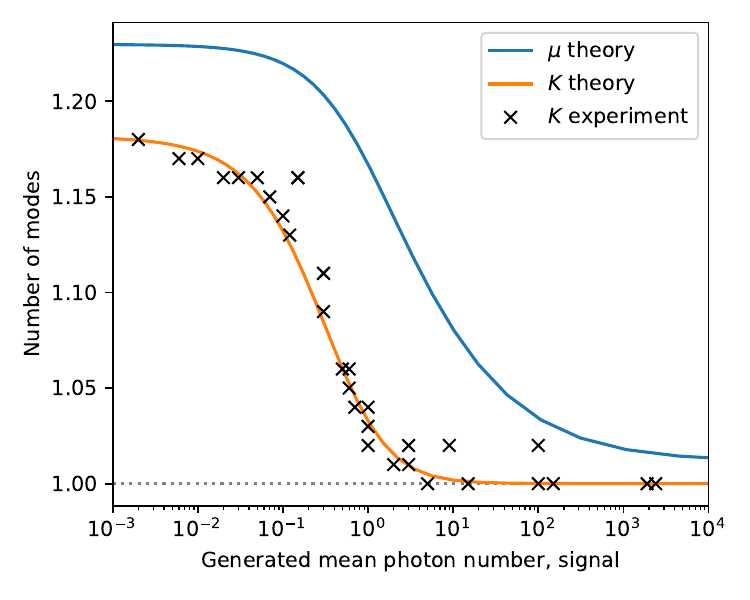}
\caption{Inferred mode number from Schmidt decomposition of the JSI. Shown are the number of modes reconstructed from JSI measurements (black crosses) and JSI simulation (orange line). For comparison, we show the number of modes predicted by the theory when one does not assume that the generated state exhibits a flat phase profile (blue line).} 
\label{fig:JSI_decomp}
\end{figure}

To estimate the number of modes from the JSI measurements, one has to make apriori assumptions about the generated state -- namely one has to assume a flat phase profile. It is known that this assumption is often invalid \cite{graffitti2018design}, but is nevertheless used due to the complexity involved in taking join spectral amplitude measurements \cite{Thekkadath2022}. To estimate the number of modes from the JSI, one  performs a Schmidt decomposition of the square root of the JSI; $\sqrt{J(\omega, \omega^\prime)} = \sum_n \sigma_n \phi_n(\omega)\psi_n(\omega^\prime)$.
The number of modes is then estimated as
 $ K = \frac{[\sum_n \sigma_k^2]^2}{\sum_n \sigma_k^4}$.
The number of modes determined through the decomposition of the measured JSI, and determined through the simulated JSI from the presented theory are shown in Fig. \ref{fig:JSI_decomp} as black crosses and an orange line, respectively.

The rigorous theory framework described in this paper also allows us to predict the true number of modes generated by the process, i.e the number of modes determined through the calculated value $\mu_a$, defined via Eq.~\eqref{eq_number_of_modes} which is not impacted by the chosen experimental measurement scheme. The predicted number of modes determined in this way is also presented in Fig. \ref{fig:JSI_decomp} (blue line). One immediately notices that the number of modes inferred from the JSI measurements, given by $K$, are significantly lower than the number of modes determined from $\mu_a$. This highlights the fact that the Schmidt decomposition of the JSI is, in general, insufficient to provide accurate information about the modal structure of the generated fields, but also shows that our modeling can accurately describe this effect.

\subsection{Spectral Broadening}

\begin{figure}
\centering
\includegraphics{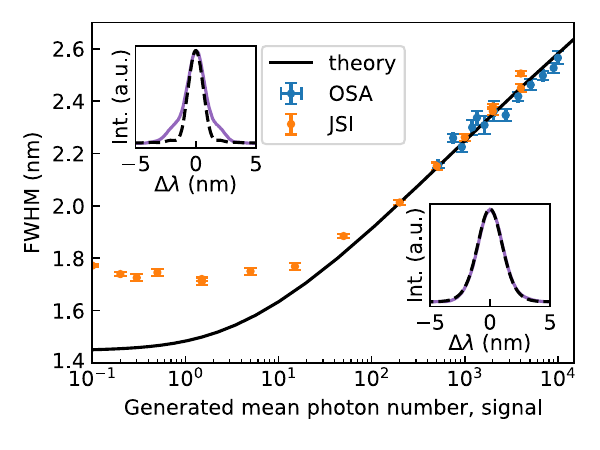}
\caption{Bandwidth versus generated mean photon number. Measured signal field bandwidth (FWHM) as a function of mean photon number from OSA measurements (blue dots), JSI measurement marginals (yellow dots) and the numerically evaluated predicted bandwidth found using Eq. \ref{eq_spectrum} (black line). Upper and lower insets show the measured (violet line) and predicted (dashed black line) centered JSI signal marginals at generated mean photon numbers of approximately 0.1 and 4000 and highlight the reason for the discrepancy between measurements and theory seen at low mean photon numbers (see main text).  
Error bars represent the standard deviation from repeated measurements (OSA) or errors in the fit (JSI).
}
\label{fig:Exp-OSA}
\end{figure}

\label{Sec:spectra}

The JSI measurements presented in the previous section reveal that the bandwidth of the generated states increases with increasing system gain. To investigate this effect further, we measure the full width at half maximum (FWHM) of the generated states by fitting Gaussian functions to the centered, central regions of the signal marginals of JSI measurements acquired at different generated mean photon numbers. The resulting FWHMs are shown in Fig. \ref{fig:Exp-OSA}. The measured FWHM values reveal a clear broadening with increasing mean photon number in strong agreement with the behavior predicted by our theory. However, a discrepancy between the measured and predicted bandwidths at lower mean photon numbers is observed, the cause of which is illustrated by the insets within the figure. The predicted (black dashed lines) and measured (violet lines) marginals at the lowest (upper inset) and highest (lower inset) generated mean photon numbers reveal a slightly distorted and widened marginal at lower gains, most likely due to waveguide inhomogeneities \cite{Fejer92,Santandrea2019}. This distortion is seen to diminish at high gain operation where the spectral bandwidth increases and the system tends towards single mode operation, as shown in Fig. \ref{fig:JSI_decomp}. 

The high brightness of the waveguided source also enables a more standard approach to spectral characterization -- at mean photon numbers close to 40,000, the pulse train contains nanowatt-scale average optical powers. The spectrum of the generated squeezed vacuum states has therefore also been measured by first separating signal and idler fields at the PBS and recording the FWHM of the signal photons using an OSA as the pump power is varied. These measurements are presented in Figure \ref{fig:Exp-OSA} and are consistent with both the FHWMs determined from JSI measurements and the theoretical predictions.

\subsection{Second-order correlation function}

\begin{figure}
\centering
\includegraphics{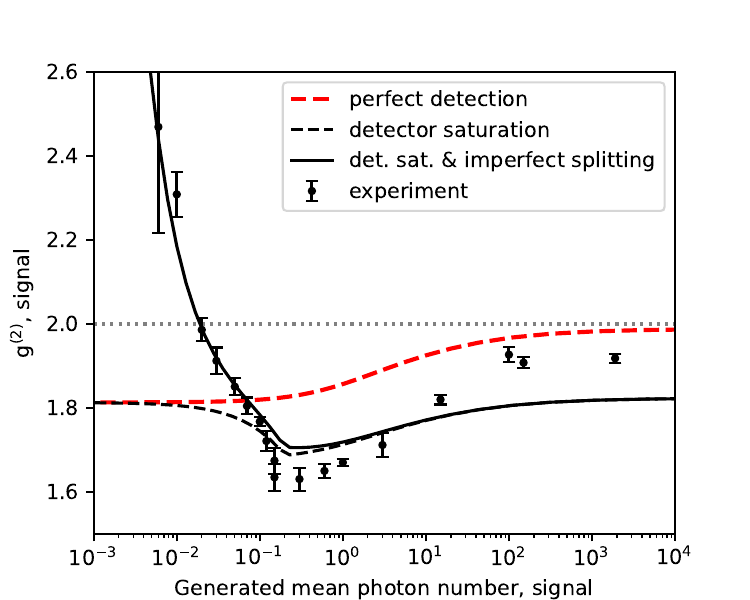}
\caption{Dependence of the $g^{(2)}(0)$ correlation function on the generated mean photon number $\langle n \rangle$ as measured via HBT inteferometry (black dots). For HBT measurements, attenuation is varied to ensure that the detected mean photon number is below 0.02 for all measurements $\langle n_{\text{det}} \rangle \leq 0.02$. Theoretical predictions for the case of perfect detection (dashed red trace), when taking into account detector saturation (dashed black trace) and when taking into account both detector saturation and imperfect PBS splitting (black trace).} Error bars represent the standard error from repeated measurements.

\label{fig:Exp-g2}
\end{figure}

In Sec. \ref{Sec:JSI}, JSI measurements were used to estimate the number of modes in the signal field. However, this approach was shown to be unreliable. The second-order correlation function, $g^{(2)}(0)$, generally provides a more robust method for estimating the number of modes. However, as we demonstrate here, this method also has its limitations.
For a perfectly single spectro-temporal mode source, the individual signal and idler fields are expected to exhibit thermal statistics, $g^{(2)}(0) = 2$ \cite{christ2011probing}, while for multimode sources, it has been shown that this value can be used to count the effective number of modes $\mu$ in the measured field $g^{(2)}(0) = 1+\frac{1}{\mu}$. 

In our experimental setup, the  $g^{(2)}(0)$ value is measured by setting the HWP to deterministically separate signal and idler and utilizing the HBT interferometer setup in the signal field as shown in Fig. \ref{fig:Exp-Setup}. 
The experimental results (black dots) are shown in Fig. \ref{fig:Exp-g2}.
It is important to note, and account for, two experimental aspects that limit the precision of the measured $g^{(2)}(0)$ values.
The first is imperfect splitting  of signal and idler fields on the PBS, an effect which is further described in Appendix \ref{App:ExpImp}.
Secondly, the limitations of click detection should be considered --particularly the impact of higher photon number events. The impact of this effect is minimized through significant attenuation for states with high photon numbers, but may still have an impact for any detected state with non-vanishing mean photon number.
Both of these effects, imperfect separation and saturation of click-detectors, are accounted for in the numerical simulations (see details in Appendix \ref{App:ExpImp}), the results of which are presented in Fig. \ref{fig:Exp-g2} (black line). 
The idealized cases of perfect signal–idler splitting (dashed black line) as well as perfect signal-idler splitting combined with unsaturated detection (dashed red line) are also presented. The model predicts a low-gain second-order correlation of 1.81, corresponding to an effective mode number of 1.23. Note that this value approaches unity with increasing gain, coinciding with the reshaping of the JSI observed in Fig. \ref{fig:Exp-JSI}. Although experimental imperfections substantially affect the measured correlation function, the model incorporating these effects remains in strong qualitative agreement with the data, validating the near-single-mode performance of the source.


\subsection{Hong-Ou-Mandel interference}

\begin{figure}
\centering
\includegraphics[]{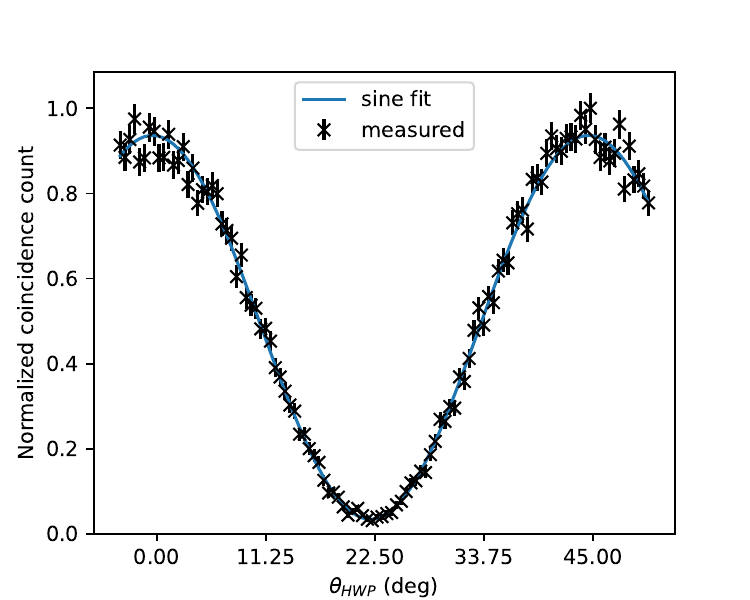}
\caption{HOM interference as polarization is varied. The measured coincidence count rate between the two outputs of the PBS as the HWP angle is varied (blue crosses). The mean number of photons in each mode is approximately 0.1. A sinusoidal fit to the data is also shown (blue line) which corresponds to a visibility of $(92.6\pm0.8)\%$.}
\label{fig:Exp-HOM}
\end{figure}

HOM interference, in which two single photons entering different  ports of a beamsplitter are observed to bunch at the output, is a commonly used technique to quantify the indistinguishability between signal and idler photons from a PDC process. In our source, indistinguishability between signal and idler fields is critical to ensure that, after interference, the generated TMSV state is faithfully transformed into two independent SMSV states.

In contrast to the standard HOM setup that varies the distinguishability between the two fields using temporal delays, we vary degree of distinguishability between the two photons using the polarization degree of freedom. In this measurement, the polarization of the signal and idler photons are rotated using the HWP before the PBS, which acts as a variable beamsplitter on both fields. The coincidence rate between the two outputs of the PBS were recorded as the angle of the HWP was varied, the results of which are shown in Fig. \ref{fig:Exp-HOM}.  
The HOM visibility for the polarization HOM dip used here is defined as $V=(0.5\cdot C_{\rm{max}}-C_{\rm{min}})/(0.5\cdot C_{\rm{max}})$, where $C_{\rm{max}}$ and $C_{\rm{min}}$ are the maximum and minimum coincidence counts, respectively. The visibility was measured in this way was found to be $V=(92.6\pm0.8)\%$, confirming a high degree of indistinguishability between signal and idler fields \cite{MeyerScott2018}.



\subsection{Number of modes in SMSV}

\begin{figure}
\centering
\includegraphics{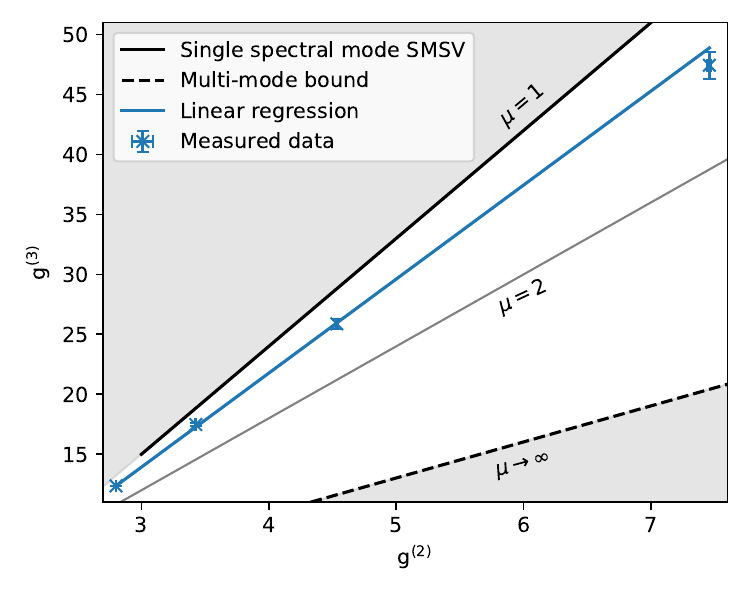}
\caption{Number of modes in the generated SMSV state. The gradient of $g^{(2)}$ vs $g^{(3)}$ is dependent on the number of modes in the SMSV state (see text). Errors shown are Poisson counting errors and are smaller than the markers for $g^{(2)}$ values below 4. Note that higher pump powers correspond to lower $g^{(2)}$ values.}
\label{fig:g2vsg3results}
\end{figure}

The number of modes contained in the SMSV states generated after interference on the PBS can be estimated using a correlation-function approach that has so far seen only limited experimental application. The method probes the relation between the pulse-integrated $g^{(2)}$ and $g^{(3)}$ correlation functions as the mean photon number of the SMSV state is varied. For a pure state with negligible ordering effects, a linear relation between these correlation functions is expected \cite{christ2011probing}, with the slope $s$ directly related to the mode number $\mu_\mathrm{SMSV}$ through $s=3 + 6/\mu_\mathrm{SMSV}$ \cite{christ2011probing,Wakui2014}.

Measuring the higher-order correlation function requires photon number resolution beyond what is shown in Fig. \ref{fig:Exp-Setup}. To this end, the fiber BS (see Fig. \ref{fig:Exp-Setup}) is replaced with an eight-bin time-multiplexed scheme, enabling pseudo photon number resolution up to 8 photons \cite{Achilles2003}. The resulting total transmission for this measurement was approximately 0.2 and the correlation functions were measured with mean photon numbers from 0.038 to 0.44, corresponding to the range over which the mode number is expected to remain roughly constant (See Fig. \ref{fig:Exp-g2}). We reconstruct the pseudo-photon-number statistics from the measured click statistics following the method presented in \cite{Krishnaswamy2024} without performing a loss inversion, as the correlation functions are insensitive to loss.

The experimentally determined correlation function values (blue crosses in Fig \ref{fig:g2vsg3results}) are calculated from the pseudo photon-number statistics $p_m$ using $g^{(n)}(0) = \frac{\sum^{8}_{m=n} p_m \frac{m!}{(m-n)!}}{\left(\sum^{8}_{m=0} p_m m\right)^n}$. Performing a linear regression through the measured points results in the solid blue line. The gradient of the linear regression ($s=8.0\pm0.4$) can be used to estimate the number of modes present in the SMSV state, resulting in a value of $K=1.24\pm 0.3$. In our theoretical approach, we determine the number of modes in the SMSV state by interfering the two modes of the generated TMSV state on a beamsplitter (See Appendix \ref{App:ExpImp}), resulting in an expected value of $\mu = 1.24$ -- in close agreement with the measurement.


\subsection{Expected Squeezing Level}
 
\begin{figure}
\centering
\includegraphics[]{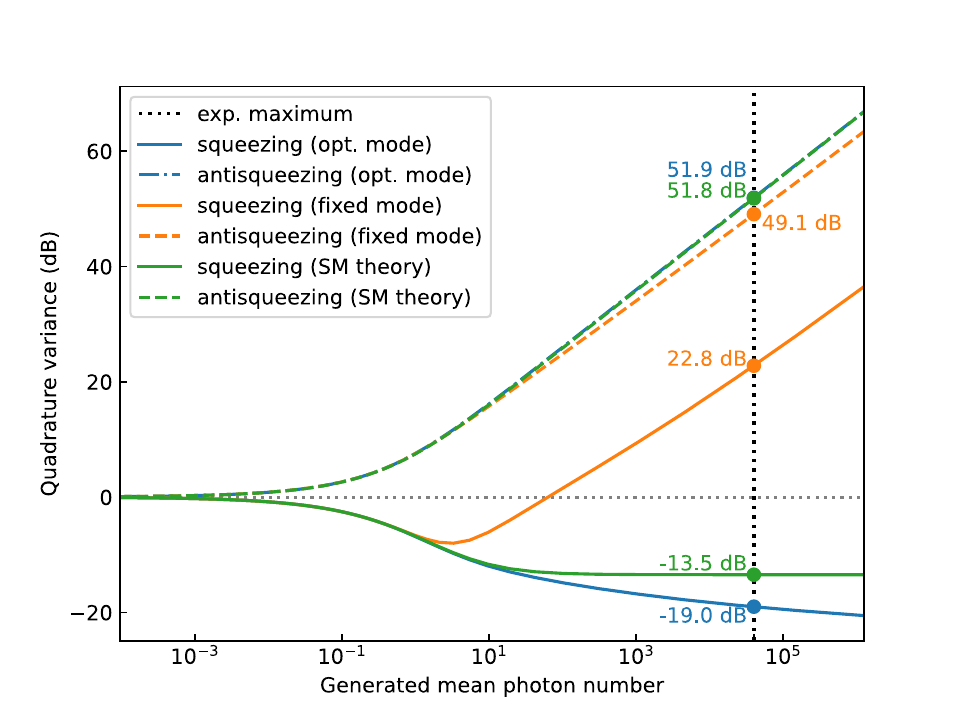}
\caption{Simulation of squeezing strength. The predicted squeezing and anti-squeezing levels produced in the KTP waveguide sample under different assumptions; single-mode theory (green trace), full theory with fixed spectro-temporal mode (orange trace) and full theory with highest squeezing found from all spectro-temporal modes (blue trace). The black dotted line indicates the maximum mean photon number measured in the experiment. Further explanation in text. }
\label{fig:sim_squeezing}
\end{figure}

In many cases, the quantum enhancement enabled by SMSV states is either proportional to the degree of squeezing available or requires a minimum squeezing level, as is the case in quantum error-correction schemes. \cite{Fukui2018,Noh2020}. It is therefore important to determine the degree of squeezing that our source is capable of producing, especially as the system is driven into the high-gain regime and the impact of both losses and ordering are no longer negligible. The naive, single-mode model that treats loss as a virtual beamsplitter after state production was presented in Section \ref{sec:Theory}, which led to a predicted squeezing level after loss given  $\text{V}_{\text{SM}} = -10 \cdot \log_{10}\left( \eta \cdot V + (1 - \eta) \right)$. The value obtained using this model will be compared to the squeezing level obtained through our rigorous model, after interfering the generated TMSV state on a beamsplitter. For both cases, the system is simulated with an internal loss corresponding to that measured in the  physical system (of 0.25 dB/cm) and we have neglected outcoupling loss due to the Fresnel reflection, as this loss can be nearly perfectly mitigated with the appropriate optical coating.

The results of the theory investigation are presented in Fig. \ref{fig:sim_squeezing}. 
The predicted squeezing and anti-squeezing levels obtained using the basic single-mode squeezing model are first shown (green trace). Using the naive, single-mode theory, the degree of squeezing predicted as the system is driven to higher gains asymptotes to a level that is determined by the internal loss of the given system (-13.5 dB). Next, we apply the rigorous theory and report the degree of squeezing found using two different treatments; in the first, the field mode where the maximum squeezing resides at low-gain (orange line) is first determined. The gain is then varied and the level of squeezing found in this particular mode is reported as the gain of the system is increased. Experimentally, this corresponds to optimizing the local oscillator of a homodyne detector at low pump powers and using the same local oscillator at high pump powers. In this case, we see that a maximum squeezing value of approximately -9 dB is would be measured at a mean photon number close to 2. Beyond this point the squeezing level is seen to decrease. Finally, we also plot the squeezing obtained when, at each mean photon number, one searches for the mode basis that optimizes the degree of squeezing. Experimentally, this corresponds to optimizing the shape and delay of the local oscillator of a homodyne detector at each mean photon number. Here, we see that the squeezing is recovered, and furthermore, it is even possible to measure more squeezing than is predicted by the basic single-mode theory. In this case, we see that the amount of squeezing produced by the source is expected to be -19.0 dB when pumping at the highest pump powers used in the experiment (as indicated by the dotted vertical line). 


\section{Discussion}

The presented comprehensive experimental and theoretical investigation into the performance of our source enables us to assess its practical suitability for hybrid quantum photonics. Furthermore, our holistic approach allows us to identify factors that limit the current performance of the device and determine whether they arise due to technical, or fundamental, limitations, allowing us to explore methods for improving upon the current system. 

(1) \textit{Scalability}: 
The source was shown to be extremely efficient, verifying its suitability for applications that require a high degree of scalability. In the low-gain regime, the source produces a mean photon number in the signal field of 20 photons per pulse per nJ of pump pulse energy. Driving the system with a pump field at 775nm and a repetition rate of 80MHz with an average power of 1mW produces 80 million -3dB squeezed states per second. On the other hand, spatial multiplexing is more difficult in this platform due to the limited reproducibility in producing rubidium exchanged waveguides in KTP, a problem which may be addressed through improved processes or alternative waveguiding methods such as laser written \cite{Chen2014} or ridge waveguides \cite{Eigner2018}.

The source was also shown to be capable of producing extremely bright single-mode SMSV states -- at the highest available pumping power of 480pJ, a generated mean photon number per pulse in the signal field of greater than 40,000 was observed. Although losses limit the degree of squeezing produced in these bright states, they nevertheless represent a useful resource for some applications. For example, it has been shown that bright downconversion sources can enhance the efficiency of nonlinear optical effects, even in the presence of significant loss \cite{spasibko2017multiphoton,Manceau2019,Rasputnyi2024,Rasputnyi2026}.

(2) \textit{Single-mode emission}: 

Single spectro-temporal mode operation of the source was verified using JSI measurements, second-order correlation measurements on the signal field, and second- and third-order correlation measurements on the SMSV state after interference on the PBS. Our theory was able to describe the measurements that were taken, even in the presence of experimental imperfections and limitations that were shown to have a large impact on the JSI and second-order correlation function measurements. The relationship between second- and third-order correlation functions was presented as an alternative approach to determine the number of modes in the generated SMSV state, revealing an effective mode number of $\mu=1.24\pm0.3$. This approach is likely more reliable than the previous methods due to the fact that the diverging correlation function values at low mean photon number represent the desired signal rather than a significant noise source. The reliability of this characterization technique in the presence of experimental imperfections is left as future work, but offers a robust scheme for modal characterization of SMSV states.

The main factor limiting single-mode operation of the presented source is the side-lobes in the phasematching spectrum arising from the sinc squared phasematching function. Tight filtering can decrease the number of modes by rejecting these side lobes, but this technique introduces loss and is ultimately limited by the incoherent filtering process \cite{MeyerScott2017}.  A more elegant method is to suppress these side lobes using apodized poling \cite{Dosseva2016,graffitti2018design,Pickston2021}. However, the cost of using this technique is a significant reduction in the effective interaction length of the sample---typically by a factor of 4 to 5 \cite{Pickston2021}. This reduction in the effective length increases the spectral bandwidth of the generated photons by the same factor -- thereby increasing the impact of fiber dispersion. As longer rubidium exchanged KTP waveguides are currently not commercially available, standard poling was favored to maintain the desired spectral bandwidth of the down-converted fields.

Finally, our investigations into the high-gain regime reveal that the system is driven towards single-mode operation at these conditions. This result is qualitatively consistent with previous investigations showing a similar evolution towards single-mode operation \cite{Sharapova2020,kumar2026}. This effect does provide a possible technique for pushing the system towards perfect single-mode operation, but the impact of losses, the resulting purity of the quantum state, and possible introduction of third-order nonlinearities at these powers need to be carefully considered ~\cite{quesada2022beyond} and require additional in-depth studies.

(3) \textit{Strong, pure squeezing}: 

We investigated the predicted squeezing performance of the sample and found that, taking into account the current waveguide loss and pump power limitations, it could theoretically produce nearly 20\,dB of squeezing. Verifying the degree of squeezing in the high-gain regime would require homodyne detection. Homodyne measurements on a similar source in the low-gain regime have been demonstrated and agree with theoretical predictions \cite{Dirmeier2020}, however, the results shown in Fig.~\ref{fig:sim_squeezing} indicate that measurements in the high-gain regime are more challenging: the spectral and temporal profile of the local oscillator must be optimized for each operating power. We leave this investigation to future work.

Some applications in hybrid quantum photonics require strongly squeezed states that also exhibit a high degree of state purity. For example, Gottesman-Kitaev-Preskill (GKP) codes, in the best case, require at least 10 dB of pure squeezing to attain error correction thresholds \cite{Fukui2018,Noh2020} and squeezed light enhancement in gravitational-wave interferometers may be limited by the coupling of anti-squeezing into the squeezing quadrature via phase fluctuations \cite{Dwyer2013}.

In our source, as is the case in general, the main factor that ultimately limits both squeezing magnitude and purity is waveguide loss. The waveguide loss in the presented system is 0.25 dB/cm, which for rubidium exchanged KTP waveguides is state-of-the-art. Although KTP is used due to its unique dispersion properties, it is a notoriously difficult material platform for producing waveguides due to effects such as ionic conductivity that render the production of periodically poled waveguides complex \cite{Padberg2022}. Reducing these losses requires improvements in material processing, or alternative architectures.

One such architecture, that provides a promising solution to the problem, is nanophotonics, such as the thin film lithium niobate platform. In these system, the optical field is tightly confined and therefore the dispersion is impacted significantly by the waveguide geometry, allowing one to tailor the physical system for the desired nonlinear process. Furthermore, in these platforms extremely low losses have been demonstrated \cite{Zhang2017}. The main issue in these systems, currently, is achieving efficient out-coupling from the tightly guiding platform \cite{labbe2025thin}.

Our in-depth investigations also revealed that ordering effects may lead to an increase in the produced degree of squeezing than would be predicted by the naive single-mode model (See Fig. \ref{fig:sim_squeezing}). This can be understood, by considering that ordering takes into account causality during the photon generation process. This causes the majority of the nonlinear process to occur at later stages through the sample, effectively reducing the losses experienced by the generated state. It is for this same reason that the squeezing level is seen to continuously improve as the system is driven harder and not saturate as is predicted by the simplified theory. We furthermore note that this explanation is consistent with the state broadening seen in Fig. \ref{fig:Exp-JSI} and \ref{fig:Exp-OSA}, as shorter interaction lengths are also be expected to increase the marginals of the JSI.

(4) \textit{Compatibility with detection schemes and networking infrastructure}: 

Our source generates squeezing centered at 1547nm -- compatible with standard fiber networks and SNSPDs, and, in the low-gain regime, produces photons with a bandwidth of approximately 1.5~nm, corresponding to a Fourier-limited pulse duration of roughly 2~ps. The bandwidth of the source was selected as the optimal compromise between a number of competing factors. The first consideration is fiber dispersion, which favors narrower bandwidths. The presented system is suitable for systems with differential fiber lengths of many tens of meters -- with fiber lengths beyond this requiring dispersion compensation or longer waveguides. Intrinsic PNR in SNSPDs instead requires pulses to be no longer than a few tens of picoseconds \cite{Schapeler2024}. Finally, faster experimental shot rates favor shorter pulses. Therefore, the optimal pulse duration to minimize fiber dispersion while still permitting intrinsic PNR in SNSPDs is something in the range of a few picoseconds to a few tens of picoseconds. Longer KTP waveguides could, in principle, be used to decrease the impact of fiber dispersion but high-quality KTP waveguides longer than 2 cm have only recently become commercially available.

\section{Conclusions and Outlook}
In conclusion, we have presented and comprehensively characterized an integrated source based on a single-pass Type-II PPKTP waveguide that is optimized for use in hybrid quantum photonics applications. By interfering the (nearly perfectly spectrally and spatially indistinguishable) generated signal and idler fields, we produce nearly perfectly single-mode SMSV states with a central wavelength at 1547 nm and picosecond duration. These pulse durations enable intrinsic PNR in SNSPDs, while minimizing the impact of fiber dispersion and avoiding limitations on experimental shot-rate. The central wavelength falls within the telecom C-band, enabling straightforward integration into existing short-reach fiber networks, as well as standard integrated photonic platforms. Furthermore, the high source efficiency enables massive temporal multiplexing in order to produce large numbers of squeezed states. 

We developed a theory that includes the impact of losses during state generation and ordering to obtain a more comprehensive understanding of the current performance and ultimate limitations of the system. A comprehensive suite of photon counting measurements, including generated mean photon number, second- and third-order correlation functions, JSIs and spectral bandwidth, were used to verify both the reliability of the model and the performance of the source. Furthermore, the predicted level of squeezing produced by the source was investigated using this theoretical framework, revealing that the platform is capable of producing squeezing levels approaching -20 dB and that driving the system in the high-gain regime pushes the system nearer to true single-mode operation. These results demonstrate that the source is readily deployable in a broad range of hybrid quantum photonics applications and motivates further investigation into understanding the fundamental performance limits of single-mode SMSV state generation.

It remains an open task to similarly verify the performance of alternative architectures under lossy, high-gain operation, although investigations in this direction have begun. Single-mode SMSV states produced using type-0 processes inside resonators have been employed in hybrid quantum photonics experiments \cite{madsen2022quantum}, and the impact of high-gain operation \cite{Vendromin2024} and lossy, high-gain operation \cite{Kim2025} in certain platforms has been theoretically investigated. In addition, high levels of squeezing (12 dB) have been demonstrated in a waveguided, continuous-wave-pumped, single-pass type-0 system \cite{ha2026generation}. Although this platform intrinsically produces highly spectrally multimode SMSV states, it can be used in hybrid quantum photonics with advanced filtering techniques \cite{Yoshikawa2017,tomoda2026} or in wholly continuous-variable schemes \cite{yokoyama2026,Larsen2019}. Of particular interest, recent results have shown that the spatial mode of the generated squeezed-light field is altered when operating in the high-gain regime \cite{ha2026generation}, naturally motivating a more rigorous theoretical treatment of this platform under high-gain operation. We therefore aim to motivate further exploration of lossy, high-gain operation in alternative architectures, in order to better understand their current capabilities and fundamental limitations. This understanding will be critical for building the next generation of quantum photonics experiments, which will place increasingly stringent demands on squeezed-light sources.

\section*{Data availability}
The experimental data that support the findings of this study are available from the corresponding author upon reasonable request.

\section*{Acknowledgments}
This work has received funding from the German Federal Ministry of Research, Technology and Space within the PhoQuant project (Grant No. 13N16103).

F.S. is part of the Max Planck School of Photonics supported by the Dieter Schwarz Foundation, the German Federal Ministry of Research, Technology and Space (BMFTR), and the Max Planck Society.


\section*{Competing interests}
The authors declare no competing interests. 

\section*{Additional information}
\textbf{Correspondence and requests for materials} should be addressed to K.-H.L. or M.S.

\appendix
\renewcommand{\thesection}{\Alph{section}}

\section{Phase matching profile}

\label{App:Phasematching}

\begin{figure}
 \centering
 \includegraphics[scale=0.8]{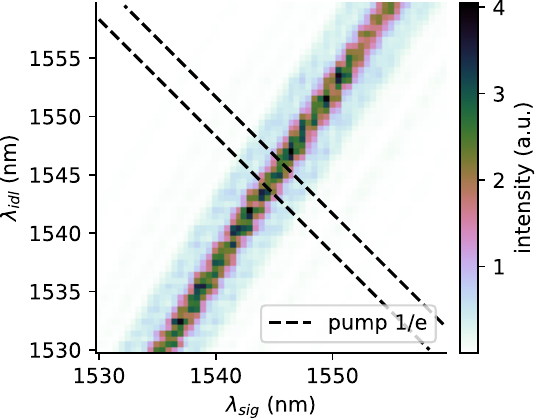}
 \caption{Measured phasematching profile. Sum frequency generation reveals a phase-matching angle of approximately 57 degrees at 1545nm.}
 \label{fig:Exp-PM}
 \end{figure}

    The phase-matching profile was experimentally characterized using sum frequency generation (SFG) from two tunable continuous-wave (CW) lasers in the telecom wavelength. 
    The measured profile is shown in Fig.~\ref{fig:Exp-PM}; the scanning was taken from 1510~nm to 1560~nm with a step size of 0.5~nm.
    The phase-matching angle is found to be approximately 57 degrees at 1545 nm.
    The experimental phase-matching was approximated with the linear phase-mismatch function
    \begin{equation}
        \Delta k(\Delta \omega, \Delta \omega^\prime) = a \Delta \omega + b\Delta \omega^\prime,
    \end{equation} 
    with the following parameters $a= 0.1709$ mm$\cdot$ps and $b=-0.1101$ mm$\cdot$ps.
    These parameters were used in numerical simulations of PDC.

\section{Detailed model for determining $g^{(2)}$ with experimental imperfections}

\label{App:ExpImp}

The implementation of the $g^{(2)}$ measurement into the numerical simulation requires careful treatment of three different effects: 1) the simulation of imperfect PBS, 2) the controlled intensity attenuation, 3) the coincidence scheme with click-detection. 

\textit{Imperfect splitting at PBS}: 

The spectrally indistinguishable, polarization-separable signal and idler fields can be interfered or separated using a PBS. Theoretically, the signal and idler fields can be perfectly separated; however, there are multiple effects that limit the separation in a real experiment. First, imperfect alignment with the PBS axis can cause incomplete signal–idler separation and consequent interference, which can generally be mitigated by rotating the polarization basis.
Second, any polarization splitter has a nonzero extinction ratio $R_E$, i.e., a small part of the idler field (here, vertical polarization) is present in the horizontal output channel of the PBS and vice versa. This effect can, in principle, be reduced through additional stages of polarization filtering. Finally, the signal and idler fields exiting the waveguide are not expected to be fully polarization orthogonal -- the waveguide modes are not pure TE- and TM-modes --  and therefore some small portion of the signal field will exit the idler port in the same polarization. This effect is stronger in more tightly confined waveguides systems \cite{babel2023}, and is typically negligible in weakly guided systems such as that presented here. However, although this effect is minor and the relative power of the SMSV state in each output mode is low, the value of the second-order correlation function diverges at low mean photon number, $g^{(2)}_{SMSV}(0) = 3 + \frac{1}{\langle n \rangle}$ \cite{Loudon2000quantum}. Therefore, even though only a very minor component of SMSV state exits each output of the PBS, this component is expected to have a large impact on correlation function measurements in the low-gain regime.
To quantitatively predict the measured second-order correlation function in the presence of this imperfection, the model must account for its effects.

To simplify the theoretical treatment, the origin of imperfect splitting at PBS is considered to originate entirely due to a limited extinction ratio. The PBS is then modeled as a lossless linear optical element with four input modes $(\hat{a}_1,\hat{b}_1,\hat{a}_2,\hat{b}_2)^T$, where modes $\hat{a}_i$ and $\hat{b}_i$ denote horizontal and vertical polarizations, respectively, and the index $i \in \set{1,2}$ represents the spatial modes. In what follows, we are interested in the field, exiting the first spatial channel of PBS. The transformations for the second spatial channel can be done similarly.

The generated PDC light, described by the matrices $\mathcal{D}_a$, $\mathcal{D}_b$ and $\mathcal{C}_{ab}$ passes through the imperfect PBS and at the first output channel the correlation matrices reveal the following transformations
\begin{equation}
   \mathcal{D}_a \rightarrow  \mathbf{t}_N \mathcal{D}_a \mathbf{t}_N, ~ ~ 
    \mathcal{D}_b \rightarrow \mathbf{r}_N \mathcal{D}_b \mathbf{r}_N , ~ ~
    \mathcal{C}_{ab} \rightarrow \mathbf{t}_N \mathcal{C}_{ab} \mathbf{r}_N ,
\end{equation}
where $\mathbf{t}_N = t \mathbf{1}_N$, $\mathbf{r}_N = r \mathbf{1}_N$,  $t=\sqrt{1-R_E}$ and $r = \sqrt{R_E}$ and $\mathbf{1}_N$ is the identity matrix.

\textit{Attenuation}: 
The second-order correlation function is typically measured using a HBT interferometer with click detectors. Reliable measurements require vanishingly small mean photon numbers, but should also be bright enough to minimize the impact of dark counts \cite{Agafonov2011}. Therefore, in order to measure states of light containing a very large number of photons per mode with click-detectors, it is necessary to attenuate the field and, ideally, take into account the impact of higher photon number events. This attenuation has been implemented using a programmable, variable attenuator, as shown in Fig. \ref{fig:Exp-Setup}. The effect of finite attenuation has been accounted for in the theory by attenuating the generated field and through thorough modeling of the click detection scheme. 
In simulations, the attenuation is realized by the transmission coefficient $T_{att}$; the correlation matrices are transformed as 
\begin{equation}
    \mathcal{D}_a \rightarrow T_{att} \mathcal{D}_a , ~ ~ 
    \mathcal{D}_b \rightarrow T_{att} \mathcal{D}_b , ~ ~ 
    \mathcal{C}_{ab} \rightarrow T_{att} \mathcal{C}_{ab} , ~ ~ 
\end{equation}

\textit{Click Detection}:
To simulate the coincidence scheme, we first separate the obtained light into two beams with a 50:50 BS.
To include the effect of imperfect splitting, we have to consider contributions from both polarizations -- i.e. our beamsplitter is described by a $4N\times4N$ unitary matrix 
\begin{equation}
    U_{BS} =       \begin{pmatrix}
                              \frac{1}{\sqrt{2}}\mathbf{1}_N & 0 & \frac{1}{\sqrt{2}}\mathbf{1}_N & 0 \\
                              0 & \frac{1}{\sqrt{2}}\mathbf{1}_N  &0 & \frac{1}{\sqrt{2}}\mathbf{1}_N  \\
                             -\frac{1}{\sqrt{2}}\mathbf{1}_N & 0 & \frac{1}{\sqrt{2}}\mathbf{1}_N & 0 \\
                              0 & -\frac{1}{\sqrt{2}}\mathbf{1}_N & 0 & \frac{1}{\sqrt{2}}\mathbf{1}_N \\
                         \end{pmatrix}.
\end{equation}

The output correlation matrices $ \mathcal{D}_{12} $ and $\mathcal{C}_{12}$ have the form
\begin{equation}
         \mathcal{D}_{12} 
         \equiv  
             \begin{pmatrix} 
              \mathcal{D}_1 & \mathcal{F} & \\
              \mathcal{F}^{H} & \mathcal{D}_2 & \\
            \end{pmatrix}  
        =
            U_{BS}^* \       
                \begin{pmatrix}
                        \mathcal{D}_a  & 0 & 0 & 0 \\
                        0 & \mathcal{D}_b  & 0 & 0 \\
                        0 & 0 & 0 & 0 \\
                        0 & 0 & 0 & 0 \\
                \end{pmatrix}  \ U_{BS}^T, 
\end{equation}

\begin{equation}
         \mathcal{C}_{12} 
         \equiv  
             \begin{pmatrix} 
              \mathcal{C}_1 & \mathcal{E} & \\
              \mathcal{E}^T & \mathcal{C}_2 & \\
            \end{pmatrix}  
        =
         U_{BS}  \ \begin{pmatrix}
          0 & \mathcal{C}_{ab}  & 0 & 0 \\
          \mathcal{C}_{ba\textbf{}} & 0 & 0 & 0 \\
          0 & 0 & 0 & 0 \\
          0 & 0 & 0 & 0 \\
     \end{pmatrix}  \ U_{BS}^T .
\end{equation}
  
The matrices $\mathcal{D}_{12}$ and $\mathcal{C}_{12}$ describe the full state after the beamsplitter, while matrices $\mathcal{D}_1$, $\mathcal{C}_1$ and $\mathcal{D}_2$, $\mathcal{C}_2$ describe the quantum state in the first and second beamsplitter output modes, respectively.
Alternatively, the full state can be described via the covariance matrix $\sigma_{12} \equiv \sigma_{12}(\mathcal{D}_{12}, \mathcal{C}_{12})$ and subsystems in first and second output modes via the covariance matrices $\sigma_{i} \equiv \sigma_{i}(\mathcal{D}_{i}, \mathcal{C}_{i})$. 

As detectors, we use two frequency-unresolved polarization-unresolved click-detectors in each channel and assume that the detection efficiency is equal for both polarizations.
The $\bar{g}^{(2)}$ measured by these detectors is then
\begin{equation}
    \bar{g}^{(2)}  = \dfrac{P_{12}}{P_{1} P_{2}},
    \label{eq:g2_prob}
\end{equation}
where the $P_{1,2}$ are the single-click probability for each of the detector, and $P_{12}$ is a coincidence probability.
The probabilities $P_{i} = 1-F(\sigma_i)$ and $P_{12} = 1 + F(\sigma_{12}) -   F(\sigma_{1}) -  F(\sigma_{2})$,
with 
$
    F(\sigma) = 2^{2N}[\sqrt{\mathrm{det}(\sigma+\mathbf{1}_{4N})}]^{-1}.
$
 For further details see Ref.~\cite{kopylov2025spectral}.

\textbf{Single-mode squeezing reconstruction:}
    To get the single-mode squeezing from the type-II PDC source, we interfere the signal and idler fields on a beasmplitter.
    The transformation of the correlation matrices is then
\begin{equation}
          \begin{pmatrix} 
              \mathcal{D}_c(\tau) & \mathcal{F}(\tau) & \\
              \mathcal{F}^H(\tau) & \mathcal{D}_d(\tau) & \\
            \end{pmatrix}
         \equiv  
             U_{BS}^*(\tau) \ \begin{pmatrix} 
              \mathcal{D}_a & 0 & \\
              0 & \mathcal{D}_b & \\
            \end{pmatrix} \ U_{BS}^T(\tau) 
\end{equation}
and 
\begin{equation}
             \begin{pmatrix} 
              \mathcal{C}_c(\tau) & \mathcal{E}(\tau) & \\
              \mathcal{E}^T(\tau) & \mathcal{C}_d(\tau) & \\
            \end{pmatrix}  
        =
         U_{BS}(\tau)  \ \begin{pmatrix}
          0 & \mathcal{C}_{ab} \\
          \mathcal{C}_{ba} & 0  
     \end{pmatrix}  \ U_{BS}^T(\tau) .
\end{equation}
with the beamsplitter transformation
\begin{equation}
    U_{BS}(\tau) =       \begin{pmatrix}
                              \frac{1}{\sqrt{2}}\mathbf{1}_N & \frac{1}{\sqrt{2}}\mathbf{1}_N \\
                            -\frac{1}{\sqrt{2}}\mathbf{1}_N & \frac{1}{\sqrt{2}}\mathbf{1}_N \\  
                         \end{pmatrix}
                         \begin{pmatrix}
                            \mathbf{1}_N & 0 \\
                            0 & \Phi(\tau) \\  
                         \end{pmatrix}
                         .
\end{equation}
The additional matrix $\Phi(\tau) = \textrm{diag}( e^{i\omega_1 \tau}, \dots ,  e^{i\omega_N \tau})$ allows us to control the time delay between signal and idler fields. 

The field at the output of the first channel is described by the correlation matrices $\mathcal{D}_c(\tau)$ and $\mathcal{C}_c(\tau)$ and therefore the covariance matrix $\sigma(\mathcal{D}_c(\tau), \mathcal{C}_c(\tau))$ can be found.
The minimal quadrature variance was found numerically via the optimization of the parameter $\tau$ and finding the the minimal-squeezing basis was found~\cite{Kopylov2025quantum}.


%

\end{document}